\documentclass{PoS}

\usepackage{amsmath} 
\usepackage{axodraw2}

\newcommand{\ba}{\begin{eqnarray}}
\newcommand{\ea}{\end{eqnarray}}
\newcommand{\fig}{figure~}
\newcommand{\eq}{eq.~}

\newcommand{\nr}[1]{(\ref{#1})}
\newcommand{\nn}{\nonumber}
\newcommand{\ep}{\epsilon}
\newcommand{\Nf}{N_{\mathrm{f}}}
\newcommand{\code}[1]{{\tt #1}}
\newcommand{\half}{\mbox{\small{$\frac{1}{2}$}}} 
\newcommand{\threehalf}{\mbox{\small{$\frac{3}{2}$}}} 
\newcommand{\threequarter}{\mbox{\small{$\frac{3}{4}$}}} 

\newcommand{\sbx}{\scalebox{0.85}}
\newcommand{\defDiag}[2]{\expandafter\newcommand%
  \csname diag-#1\endcsname{#2}}
\newcommand{\diag}[1]{\csname diag-#1\endcsname}

\newcommand{\TLfig}[1]{{\begin{array}{c}\diag{#1}\\[2ex]\text{\small #1}\end{array}}}

\newcommand{\picj}[1]{\;\parbox[c]{40pt}{\begin{picture}(40,40)(0,0)
\SetWidth{1.0}\SetScale{1.0} #1 \end{picture}}\;}
\newcommand{\picjj}[1]{\parbox[c]{0pt}{\begin{picture}(0,40)(43,0)
\SetWidth{1.0}\SetScale{1.0} #1 \end{picture}}}
\newcommand{\picbj}[1]{\;\parbox[c]{60pt}{\begin{picture}(60,40)(0,0)
\SetWidth{1.0}\SetScale{1.0} #1 \end{picture}}\;}

\def\Asc(#1,#2)(#3,#4,#5){\CArc(#1,#2)(#3,#4,#5)}
\def\Lsc(#1,#2)(#3,#4){\Line(#1,#2)(#3,#4)}

\defDiag{J}{\sbx{\picj{
\Asc(20,20)(20,-90,270)}}}

\defDiag{7}{\sbx{\picj{
\Asc(20,20)(20,0,180) 
\Asc(20,20)(20,180,360)
\Lsc(40,20)(0,20)}}}

\defDiag{51}{\sbx{\picj{
\Asc(20,20)(20,0,180)
\Asc(20,2)(27,42,138) 
\Asc(20,38)(27,222,318)
\Asc(20,20)(20,-180,0)}}}

\defDiag{63}{\sbx{\picj{
\Asc(20,20)(20,-30,90)
\Asc(20,20)(20,90,210)
\Asc(20,20)(20,210,330)
\Lsc(3,10)(20,20) 
\Lsc(20,20)(20,40) 
\Lsc(37,10)(20,20)}}}

\defDiag{841}{\sbx{\picj{
\Asc(20,20)(20,0,180)
\Asc(20,30)(22.36,-153.43,-26.57)
\Lsc(0,20)(40,20)
\Asc(20,10)(22.36,26.57,153.43)
\Asc(20,20)(20,-180,0)}}}

\defDiag{1011}{\sbx{\picj{
\Asc(20,20)(20,90,180)
\Asc(20,20)(20,180,270) 
\Asc(20,20)(20,270,360) 
\Asc(20,20)(20,0,90)
\Lsc(20,20)(20,40) 
\Lsc(20,20)(20,0) 
\Lsc(0,20)(20,20) 
\Lsc(40,20)(20,20)}}}

\def\JSe(#1,#2,#3,#4,#5,#6){\picj{
#6(20,20)(20,0,180)
#2(20,30)(22.36,-153.43,-26.57)
#3(20,50)(36.05,-122.69,-57.31)
#4(20,-10)(36.05,57.31,122.69)
#5(20,10)(22.36,26.57,153.43)
#1(20,20)(20,-180,0)}}
\defDiag{28686}{\sbx{\JSe(\Asc,\Asc,\Asc,\Asc,\Asc,\Asc)}}

\def\JSy(#1,#2,#3,#4,#5,#6,#7,#8){\picj{
#1(20,20)(20,0,90)
#2(20,20)(20,-180.,-90)
#3(40,0)(20,90,180)
#4(0,40)(20,-90,0)
#5(40,40)(20,-180,-90)
#6(0,0)(20,0,90)
#7(20,20)(20,-90.,0)
#8(20,20)(20,90,180)}}
\defDiag{30876}{\sbx{\JSy(\Asc,\Asc,\Asc,\Asc,\Asc,\Asc,\Asc,\Asc)}}

\def\JSV(#1,#2,#3,#4,#5,#6,#7,#8,#9){\picj{
#1(20,20)(20,72,144)
#2(20,20)(20,144,216)
#3(20,20)(20,0,72)
#4(20,20)(20,-144,-72)
#5(20,20)(20,-72,0)
#6(3.81966, 31.7557)(20,20)
#7(26.1803, 39.0211)(20,20)
#8(3.81966, 8.24429)(20,20)
#9(40,20)(20,20)}}
\def\JSVt(#1){\picjj{
#1(26.1803, 0.97887)(20,20)}}
\defDiag{32596}{\sbx{\JSV(\Asc,\Asc,\Asc,\Asc,\Asc,\Lsc,\Lsc,\Lsc,\Lsc)\JSVt(\Lsc)}}

\defDiag{32279}{\sbx{\picbj{
\Lsc(20,0)(40,0)
\Lsc(20,40)(40,40)
\Asc(20,20)(20,-180,-90)
\Asc(40,20)(20,-90,90)
\Lsc(20,20)(40,20)
\Lsc(20,0)(20,20)
\Lsc(40,20)(40,0)
\Lsc(20,20)(20,40)
\Lsc(40,20)(40,40)
\Asc(20,20)(20,90,180)
\Lsc(20,0)(40,20)}}}

\defDiag{32745}{\sbx{\picbj{
\Lsc(20,0)(40,0)
\Lsc(20,40)(40,40)
\Asc(20,20)(20,-180,-90)
\Asc(40,20)(20,-90,90)
\Lsc(20,20)(40,20)
\Lsc(20,0)(20,20)
\Lsc(40,20)(40,0)
\Lsc(20,20)(20,40)
\Lsc(40,20)(40,40)
\Asc(20,20)(20,90,180)
\Lsc(30,0)(30,20)}}}

\defDiag{31740}{\sbx{\picbj{
\Lsc(20,0)(40,0)
\Lsc(20,40)(40,40)
\Asc(20,20)(20,-180,-90)
\Asc(40,20)(20,-90,90)
\Lsc(20,0)(20,20)
\Lsc(40,20)(40,0)
\Lsc(20,20)(20,40)
\Lsc(40,20)(40,40)
\Asc(20,20)(20,90,180)
\Lsc(20,13)(40,13)
\Lsc(20,27)(40,27)}}}

\defDiag{30527}{\sbx{\picbj{
\Asc(20,20)(20,90,270)
\Asc(40,20)(20,-90,90) 
\Lsc(40,40)(20,40) 
\Lsc(20,0)(40,0) 
\Lsc(20,0)(20,20)
\Lsc(20,20)(40,40) 
\Lsc(20,40)(28,33) 
\Lsc(33,28)(40,20) 
\Lsc(40,20)(40,0)
\Lsc(20,20)(40,20)
\Lsc(30,0)(30,20)}}}

\defDiag{30699}{\sbx{\picbj{
\Asc(20,20)(20,90,270)
\Asc(40,20)(20,-90,90) 
\Lsc(40,40)(20,40) 
\Lsc(20,0)(40,0) 
\Lsc(20,0)(20,40)
\Lsc(40,40)(40,0) 
\Lsc(20,15)(40,15) 
\Lsc(30,12)(30,0)
\Asc(20,15)(10,15,90)}}}

\newcommand{\figur}{%
\begin{figure}[t]
\begin{center}
\begin{align*}
\begin{array}{l@{\hspace{5mm}}ccccccc}
&\TLfig{J} 
&\TLfig{7} 
&\TLfig{51} 
&\TLfig{63}
&\TLfig{841}
&\TLfig{1011} 
&\TLfig{28686} 
\\[7mm]
&\TLfig{30876} 
&\TLfig{32596} 
&\TLfig{32279} 
&\TLfig{32745} 
&\TLfig{31740} 
&\TLfig{30699} 
&\TLfig{30527} 
\\[-4mm]
\end{array}
\end{align*}
\end{center}
\caption{\label{fig}Momentum integrals $I_n$ discussed here. 
Solid lines denote massive propagators $1/(k^2+m^2)$.}
\end{figure}}

\title{Five-loop massive tadpoles}

\ShortTitle{Five-loop massive tadpoles}

\author{Thomas Luthe\\Institut f\"ur Theoretische Teilchenphysik, Karlsruhe Institute of Technology, Karlsruhe, Germany\\E-mail: \email{thomas.luthe@kit.edu}}
\author{\speaker{York Schr\"oder}\\Grupo de Fisica de Altas Energias, Universidad del Bio-Bio, Casilla 447, Chillan, Chile\\E-mail: \email{yschroeder@ubiobio.cl}}

\abstract{We provide an update on a long-term project that aims at evaluating massive 
vacuum integrals at the five-loop frontier, with high precision and in various space-time 
dimensions. A number of applications are sketched, mainly concerning the determination 
of anomalous dimensions, for quantum field theories in four, three and two dimensions.}

\FullConference{Loops and Legs in Quantum Field Theory\\
   24-29 April 2016\\
    Leipzig, Germany
 \hfill
 \begin{picture}(0,0)
 \put(-47,585){\normalfont TTP16-037}
 \end{picture}
   }

\begin{document}


\section{Introduction}

In the light of ongoing large-scale experimental effort to search for new particles at the
high-energy frontier, with the daunting task of filtering signals out of the overwhelming
Standard Model (and, in particular, QCD) background processes,
the importance of evaluating the fundamental building blocks that enter high-precision
perturbative expansions cannot be overstated.
One class of such fundamental building blocks are so-called master integrals,
which arise in modern Feynman-diagrammatic multi-loop calculations.
In order to optimize the theoretical effort that is invested in obtaining the necessary 
high-precision determinations of physical observables, it is often desired to once and 
for all tabulate universal elements (such as these master integrals) that are required often, 
but are difficult or time-consuming to obtain.

Master integrals come in all sorts of flavors, and can be classified according to 
number of loops, external invariants, propagator masses, space-time dimension,
finiteness, or number content, to name a few properties. Integration being an art rather than
a mechanic task such as differentiation, only a small subset of master integrals
is expected to be amenable to analytic methods, while the majority will need to be
approximated numerically, a task complicated by ultraviolet and infrared divergences 
and -subdivergences that can be present. 

In this contribution, pushing on the loop-frontier,
we will focus on one of the simplest possible integral classes: fully massive tadpoles,
which correspond to vacuum diagrams without external lines, in which all particles/propagators 
share a common mass. 
The mass-dependence then follows trivially from dimensional arguments, 
such that,  without loss of generality, we can set the mass parameter to unity, leaving us
with a zero-scale problem of computing pure numbers.
Indeed, over the years there have already been a fair number of talks concerning this very 
class of basic integrals at this conference series, elucidating any subset of aspects listed
above. We wish to continue this discussion, adding news at five loops.


\section{Method}

As mentioned in the introduction, we focus on the specific class of fully massive tadpoles
up to 5 loops. The 5-loop integral family needs 15 propagators, of which maximally 12 can 
be present in a vacuum graph, see \fig{\ref{fig}} for examples. 
There exist a number of viable methods evaluating such integrals, such as integration in coordinate
space, differential equations (in a mass ratio, letting $M/m\rightarrow1$ in the end), 
or numerical solutions of difference equations (in a symbolic propagator power $X$ that is set to unity the end)
via factorial series. We favor the latter approach, as formulated algorithmically by 
Laporta \cite{Laporta:2000dc,Laporta:2001dd}.

Laporta's setup, based on integration by parts \cite{Chetyrkin:1981qh}, has already been 
successfully applied to the evaluation of a large number of Feynman integrals, and in 
particular also to 4-loop massive tadpoles in 3d \cite{Schroder:2003kb} 
and 4d \cite{Laporta:2002pg,Schroder:2005va}. This makes a 5-loop extension 
natural; for some first results, see \cite{Luthe:2016sya}. 
Our method has been developed and described in more detail in \cite{Moller:2012dop} and
\cite{Luthe:2015ngq}, where the latter reference in particular contains a substantial 
fine-tuning of the Laporta approach in order to delay intermediate expression swell, 
implementing ideas such as using coupled equations, reducing recurrence relations, and predicting
instability factors, and presents the program \code{TIDE}.
Regarding results given earlier \cite{Luthe:2016sya} and below, we would like to note that,
in order to obtain convention-independent results,
we choose to divide each $L$\/-loop integral by the respective power of
the 1-loop massive tadpole $J^L$ which, picking a concrete momentum integral measure
and working in Euclidean space-time, reads $J=\int\!{\rm d}^dk/(k^2+1)=\pi^{d/2}\,\Gamma(1-d/2)$.

\section{Results and checks}

\figur

We have obtained numerical results for the $\ep$\/-expansions of the set of fully massive 
tadpoles at 5 loops, most of which are new. In practice, we aimed at about 300 digits of final
precision for the first ten $\ep$\/-orders in expansions around $d_0=2,3$ and $4$ dimensions.
Our results cover the full set of 63 5-loop sectors with 5 to 11 lines, amassing a total of 103 master
integrals. A number of results for four-dimensional expansions of 5-loop masters have already 
been given elsewhere \cite{Luthe:2015ngq,Luthe:2016sya}. 
Presently, mainly due to limitations in computer resources, we are unable to obtain results
for the four 5-loop sectors with 12 lines, depicted as $I_{32745}$, $I_{31740}$, $I_{30699}$ and $I_{30527}$
in \fig\ref{fig}, which together contain a total of 9 masters integrals. 
Further optimizations of the code \code{TIDE} \cite{Luthe:2015ngq} are under way, and we are confident 
that these last four sectors will be solved in the near future. 

As has already been mentioned above, in the program \code{TIDE}, 
a number of new ideas and improvements over the original 
Laporta algorithm have been implemented. Therefore, it is important to perform cross-checks
on the results the code delivers. To this end, we have performed successful checks against
known lower-order results, in particular at four loops in 4d \cite{Laporta:2002pg,Schroder:2005va} 
as well as 3d \cite{Schroder:2003kb}. 

Furthermore, there is an important internal check that is inherent in the method of difference
equation that we are employing. Except for totally symmetric graphs (which correspond to
our 5-loop sector numbers 28686, 30876 and 31740, see \fig\ref{fig}), 
for all so-called corner integrals (those with all propagator powers equal to unity)
\code{TIDE} generates a number of independent results coming from deriving difference
equations with power $X$ on inequivalent lines. All of these results have to coincide
at $X=1$, and they do within our numerical precision for expansions around different
values (we have explicitly checked 4d, 3d, and 2d expansions) of the space-time dimension.

We can also look at specific sets of integrals that are known analytically. As an example, 
we take the 2-loop massive sunset vacuum integral, for which a $d$\/-dimensional
analytic solution in terms of hypergeometric functions is available \cite{Davydychev:1992mt,Schroder:2005va}.
The normalized 2-loop sunset integral can then be expanded to all orders in $\ep$ in various dimensions,
for example:
\ba
I_7/J^2 &\stackrel{4-2\ep}=& -\threehalf-\threehalf\,\ep+3(3 H_2-1)\,\ep^{2}
+3(3 H_2-6 H_3-2)\,\ep^{3}
+6(3 H_2-3 H_3+6 H_4-2)\,\ep^{4}
+\dots\;,\nn\\
I_7/J^2 &\stackrel{2-2\ep}=& 6 H_2 \,\ep^{2}-12 H_3 \,\ep^{3}+24 H_4\,\ep^{4}
-48 H_5 \,\ep^{5}+96 H_6\,\ep^{6} +\dots\;,
\ea
where the rather compact expressions are due to having introduced the set of transcendentals 
\ba
H_n &=& h_n + h_1 {\hat C}_{n-1}
\Big( 1 - \frac{3^{\ep/2}\Gamma(1-\ep)}{\Gamma^2(1-\ep/2)} \Big)
\;,\quad
h_n \;=\; {}_{n+1}F_n \big( \half,\ldots,\half; \threehalf,\ldots,\threehalf; \threequarter \big) \;,
\ea
where the operator ${\hat C}_{n}$ picks out the coefficient of order $\ep^n$.
For example, $H_1=2\pi/\sqrt{27}$, while $3H_2=\sqrt{3}\,{\rm Cl}_2(2\pi/3)$ 
is related to the Clausen function ${\rm Cl}_n(x)=\sum_{k>0}\sin(kx)/k^n={\rm Im}\,{\rm Li_n(e^{ix})}$  that had been 
studied in relation with massive 3-loop vacuum integrals \cite{Broadhurst:1998rz}.
We observe that the corresponding numerical results for $I_7$ delivered by \code{TIDE} 
agree perfectly with the expansions given above, in four as well as two dimensions.
Generalizing the sunset integral to $(L+1)$ lines connecting two vertices, the corresponding
integral class $S_L$ can in principle be studied numerically in any dimension and at any loop order
using coordinate-space techniques \cite{Groote:2005ay}. 
Our lower-loop results $I_7$, $I_{51}$ and $I_{841}$ compare favorably with $S_2$, $S_3$ and $S_4$, 
respectively. At five loops, the corresponding result for
$S_5/J^5$ is easily superseeded in precision by \code{TIDE}, which delivers at least 250 digits
of (we only show 50 digits here; recall that we divide by the 1-loop tadpole, which corresponds to 
multiplying 4d 5-loop integrals by $\ep^5$)
\ba
I_{28686}/J^5 &=&  
 -2.9999999999999999999999999999999999999999999999999\,\ep^{0}
 \nn\\&&-1.5000000000000000000000000000000000000000000000000\,\ep^{1}
 \nn\\&&+0.5416666666666666666666666666666666666666666666666\,\ep^{2}
 \nn\\&&-0.8798611111111111111111111111111111111111111111111\,\ep^{3}
 \nn\\&&-1.2132523148148148148148148148148148148148148148148\,\ep^{4}
 \nn\\&&+135.95072868792871461956492733702218574897992953584\,\ep^{5}
\;.
\ea

Similarly, for another check at five loops, we consider the class $W_L$ of wheel-type integrals with $L$ 
spokes. In four dimensions, their leading $\ep$\/-order is known to all loop orders \cite{Broadhurst:1985vq}
\ba
\frac{W_L}{J^L} &=& \frac{(-1)^L\,\Gamma(2L-1)}{\Gamma(L)\,\Gamma(L+1)}\,\zeta(2L-3)\,\ep^{L-1}
+{\cal O}(\ep^L)\;,
\ea
with the leading term of our result for 
\ba
I_{32596}/J^5 &=& 
 -14.116889883346919575757165697897154634398089847913\,\ep^{4}
 \nn\\&&+235.07729596783467131454388080950411779239347239580\,\ep^{5}
 \nn\\&&-2267.7386832930084122962994480580205855487545413738\,\ep^{6}
\ea
matching the five-loop case $W_5/J^5=-14\zeta(7)\ep^4$ to 250 digits, and our lower-loop results $I_{1011}$ and $I_{63}$ matching $W_4$ and $W_3$ to tens of thousands digits.
In the same spirit, considering the family $Z_L$ of $L$\/-loop zigzag-type integrals, 
we can compare with known all-loop-order results in four dimensions, which in terms of the previous class
reads \cite{Kazakov:1984km,Broadhurst:1995km,Brown:2012ia} 
\ba
\label{eq:zigzag}
\frac{Z_L}{J^L} &=& \frac{W_L}{J^L}\,\frac4L\Big(1+\frac{(-1)^L-1}{2^{2L-3}}\Big)
+{\cal O}(\ep^L)\;.
\ea
The leading term of our result for 
\ba
I_{32279}/J^5 &=& 
 -11.117050783135699165908767987094009274588495755231\,\ep^{4}
 \nn\\&&+181.78223928612340820790788236018642961198741994209\,\ep^{5}
 \nn\\&&-1725.9996137403520805951673992442117288607704957542\,\ep^{6}
\ea
agrees with $Z_5/J^5=-\frac{441}{40}\zeta(7)\ep^4$ 
predicted by \eq\nr{eq:zigzag} to our 5-loop numerical precision of 250 digits, while the lower-loop ones
agree to much higher precision (at 4 and 3 loops, the two integral families actually coincide, $W_4=Z_4$
and $W_3=Z_3$).

Another standard method to evaluate Feynman integrals is integration over the 
Feynman-parametric representation (see, e.g.\ \cite{Smirnov:2012gma}), where the integrand is given
by two characteristic polynomials (Symanzik polynomials, usually called $U$ and $F$) that are
fixed by the topology of the underlying graph.
In most practical cases, this amounts to numerical integration, and a number of
public implementations exist to facilitate this task \cite{Bogner:2007cr,Borowka:2015mxa,Smirnov:2015mct}.
We can use this method here for some low-precision checks of our results,
and to estimate the magnitude of the missing integrals.
It turns out that for the class of fully massive integrals without external legs,
the graph polynomials coincide in general (since $F=(x_1+\dots+x_N)U=U$). Therefore, 
a fully massive $L$\/-loop vacuum integral in $d$ dimensions, having $N$ propagators
with powers $a_1,\dots,a_N$, normalized to $J^L$ (where $J$ is the 1-loop massive tadpole,
containing the momentum integral measure of choice)
can be represented as an integral over the $N$\/-dimensional simplex with a particularly simple integrand,
\ba
\label{eq:simplex}
\frac{I_{\vec{a}}}{J^L} &=& \frac{\Gamma(A-Ld/2)}{[\Gamma(1-d/2)]^L}\int_0^\infty\!\!\!{\rm d}^Nx\;
\delta(1-X)\,\frac{p_{\vec{a}}(\vec{x})}{[U(\vec{x})]^{d/2}} \;.
\ea
We have denoted the sum of indices as $A=a_1+\dots+a_N$, and $X=x_1+\dots+x_N$ is
the sum of integration variables. The integrand contains a numerator 
$p_{\vec{a}}(\vec{x})=\prod_{i=1}^N x_i^{a_i-1}\!/\Gamma(a_i)$ that is different from
unity for integrals with dots, 
and the graph polynomial U that is homogeneous of degree $L$.
Typically, a so-called primary sector decomposition \cite{Roth:1996pd,Binoth:2003ak}
is performed on \eq\nr{eq:simplex},
whence the integration domain simplifies to a $(N-1)$\/-dimensional hypercube
\ba
\label{eq:hypercube}
\frac{I_{\vec{a}}}{J^L} &=& \frac{\Gamma(A-Ld/2)}{[\Gamma(1-d/2)]^L}
\int_0^1\!\!\frac{{\rm d}^{N-1}z}{(1+Z)^{A-Ld/2}}\,
\sum_{s=1}^N\frac{p_{\vec{a}}(\vec{z}_s)}{[U(\vec{z}_s)]^{d/2}} \;.
\ea
Here, $Z=z_1+\dots+z_{N-1}$ is the sum of integration variables, and 
$\vec{z}_s=(z_1,\dots,z_{s-1},1,z_s,\dots,z_{N-1})$ fits the $(N-1)$ integration variables 
into an $N$\/-vector.
All polynomials in \eq\nr{eq:hypercube} are positive semidefinite, such that we can use a 
simple numerical integration routine to verify many of our results to low precision,
and confirm that the 12-line master integrals are indeed finite in $d=4$.


\section{Applications}

As a first application of our (4d) five-loop master integrals, we have set out to determine
the set of anomalous dimensions in QCD \cite{Luthe:2016ima}. The calculation follows
standard procedures, utilizing \code{Qgraf} \cite{Nogueira:1991ex} 
for diagram generation and 
\code{FORM} \cite{Kuipers:2012rf} 
for most computer algebraic manipulations. To isolate the 
overall ultraviolet divergences of the (massless) two- and three-point functions, one 
can either re-route external momenta through the diagram in clever ways in order to 
systematically cancel infrared divergences \cite{Vladimirov:1979zm,Chetyrkin:1984xa}, 
or introduce masses into all propagators, at the expense of one additional (mass-) 
counterterm \cite{vanRitbergen:1997va,Czakon:2004bu}. 
We choose to follow the second option, and map all integrals onto fully 
massive tadpoles by recursively applying the 
identity \cite{Misiak:1994zw,Chetyrkin:1997fm} 
\ba
\frac1{(k+q)^2} &=& \frac1{k^2+m^2}\Big(1+\frac{m^2-2kq-q^2}{(k+q)^2}\Big)
\;,
\ea
where $k$ ($q$) are loop (external) momenta and where integrals that are finite by naive 
power-counting are dropped, eventually ending the recursion.
Reduction of the large number of resulting scalar integrals to master integrals is done
via integration by parts \cite{Chetyrkin:1981qh}, for which a number of public codes
are available \cite{vonManteuffel:2012np,Lee:2013mka,Smirnov:2013dia}.
However, due to the complexity at five loops, we found it necessary to use our own in-house
programs \code{crusher} \cite{crusher} and \code{TIDE} \cite{Luthe:2015ngq} for the integral reduction. 
The latter is written almost entirely 
in \code{C++}, expect that it uses \code{Fermat} \cite{fermat} for the polynomial algebra, 
and runs all time-critical code in parallel.
Along these lines, we have been able to partially generalize results for the 5-loop SU(3) 
QCD Beta function presented by the Karlsruhe group \cite{Baikov:2016tgj} to a general 
gauge group. Concretely, so far we have completely reproduced the Beta function up to four 
loops, and obtained the $\Nf^4$ and $\Nf^3$ contributions at five loops for a general
gauge group \cite{Luthe:2016ima}, confirming the corresponding $SU(3)$ results.

The ability to change the space-time dimensionality in our setup and demand $\ep$\/-expansions
around any desired integer dimension opens up a number of other fields of application (besides
the obvious, evaluating the remaining renormalization constants of QCD). In particular, there are
interesting questions related to renormalization-group functions in three-dimensional field theories, 
such as whether an analogue of Zamolodchikov's two-dimensional $c$-theorem
can be proven in odd dimensional supersymmetric theories \cite{n2susy}.
Furthermore, allowing for zero-mass propagators as well, even the constant parts of 3d expansions 
of zero-scale master integrals do contribute to physical observables, such as for example in 
QCD thermodynamics, where they enter an effective theory treatment of the strongly interacting
gluons present in the hot early-universe plasma, see e.g.\ \cite{Kajantie:2003ax}.
Going finally to two dimensions, massive tadpoles can be used to study non-trivial 
properties of QCD-like theories such as the Gross-Neveu (GN) model, where evanescent operators
spoil multiplicative renormalizability \cite{Gracey:2016mio}. The GN model is particularly interesting due to
its connection with problems in condensed-matter theory, such as the evaluation of  critical
exponents.


\section{Conclusions}

In the era of highly automated perturbative calculations, one key building
block are master integrals. We have reported progress on evaluating the set of five-loop
massive vacuum integrals, which play an important role in determining anomalous 
dimensions of various quantum field theories. 
Since we are able to obtain high-precision numerical results for the $\ep$\/-expansions 
around any number of dimensions $d=d_0-2\ep$, applications include determinations of 
(a) the QCD Beta function \cite{Luthe:2016ima} and other anomalous
dimensions at $d_0=4$;
(b) effective theory contributions to thermodynamic 
observables \cite{Kajantie:2003ax}, 
or renormalization functions of supersymmetric gauge theories \cite{n2susy} at $d_0=3$;
(c) Beta functions of QCD-like models relevant for critical exponents 
in condensed matter systems such as graphene \cite{Gracey:2016mio} at $d_0=2$.

As an open challenge, due to the complexity of the corresponding difference- and recursion 
relations, the evaluation of the missing 12-line integrals is left for future work. 
These four sectors of trivalent 5-loop graphs are finite (at $d_0=4$), however, such that
they are not expected to contribute to the anomalous dimensions. 


\section*{Acknowledgments}

Y.S.\ acknowledges support from FONDECYT grant 1151281 and UBB project GI-152609/VC.
All diagrams were drawn with Axodraw~\cite{Collins:2016aya}.


\end{document}